\documentclass[nofootinbib,aps,prl,showkeys,groupedaddress,notitlepage]{revtex4}

\usepackage{xspace}
\usepackage{amsmath,amssymb}
\usepackage{graphicx}
\usepackage[caption=false]{subfig}

\preprint{NuFact15 - Rio de Janeiro, Brazil - August 2015}

\newcommand{\senseSindrum}{\ensuremath{7\times10^{-13}}\xspace}
\newcommand{\sensePI}{\ensuremath{3\times10^{-15}}\xspace}
\newcommand{\sensePII}{\ensuremath{3\times10^{-17}}\xspace}

\newcommand{\mueconv}{\ensuremath{\mu\text{-}e}~conversion\xspace}

\newcommand{\sindrumII}{\mbox{SINDRUM-II}\xspace}
\newcommand{\phaseI}{\mbox{Phase-I}\xspace}
\newcommand{\phaseII}{\mbox{Phase-II}\xspace}
\newcommand{\alcap}{AlCap\xspace}

\newcommand {\degree}{\ensuremath{^\circ}\xspace}

\newcommand{\Fig}[1]{Fig.~\ref{fig:#1}\xspace}
\newcommand{\fig}[1]{Fig.~\ref{fig:#1}\xspace}

\newcommand{\figlabel}[1]{\label{fig:#1}\xspace}

\makeatletter
\def\l@subsubsection#1#2{}
\makeatother

\newcommand{\FigFeynmanDiagrams}[1]{
\begin{figure}[#1]
\centering
\subfloat[][\label{fig:FD-Z-h}Z-prime and extended Higgs couplings]{\includegraphics{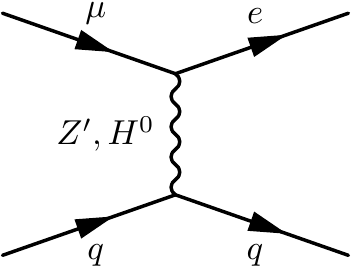}}\hspace{1cm} 
\subfloat[][\label{fig:FD-dipole}Penguin diagram with SM or heavy neutrinos, or SUSY particles]{\includegraphics{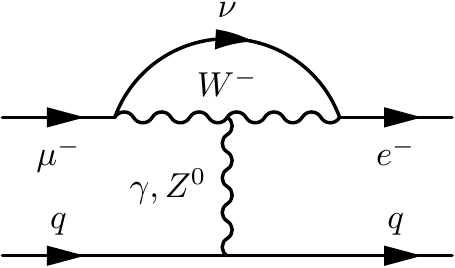}}\hspace{1cm} 
\subfloat[][\label{fig:FD-Leptoquark}Leptoquark]{\includegraphics{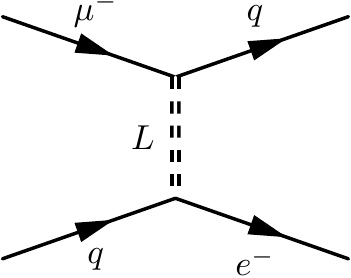}}
\caption{\label{fig:feynman_diagrams}
Example Feynman diagrams that give rise to \mueconv.
The dipole term with a SM neutrino in \protect\subref{fig:FD-dipole}, whilst allowed in the SM with neutrino oscillations, is heavily GIM suppressed.
}
\end{figure}
}

\newcommand{\FigPhaseII}[2]{
\begin{figure}[#1]
\includegraphics[#2]{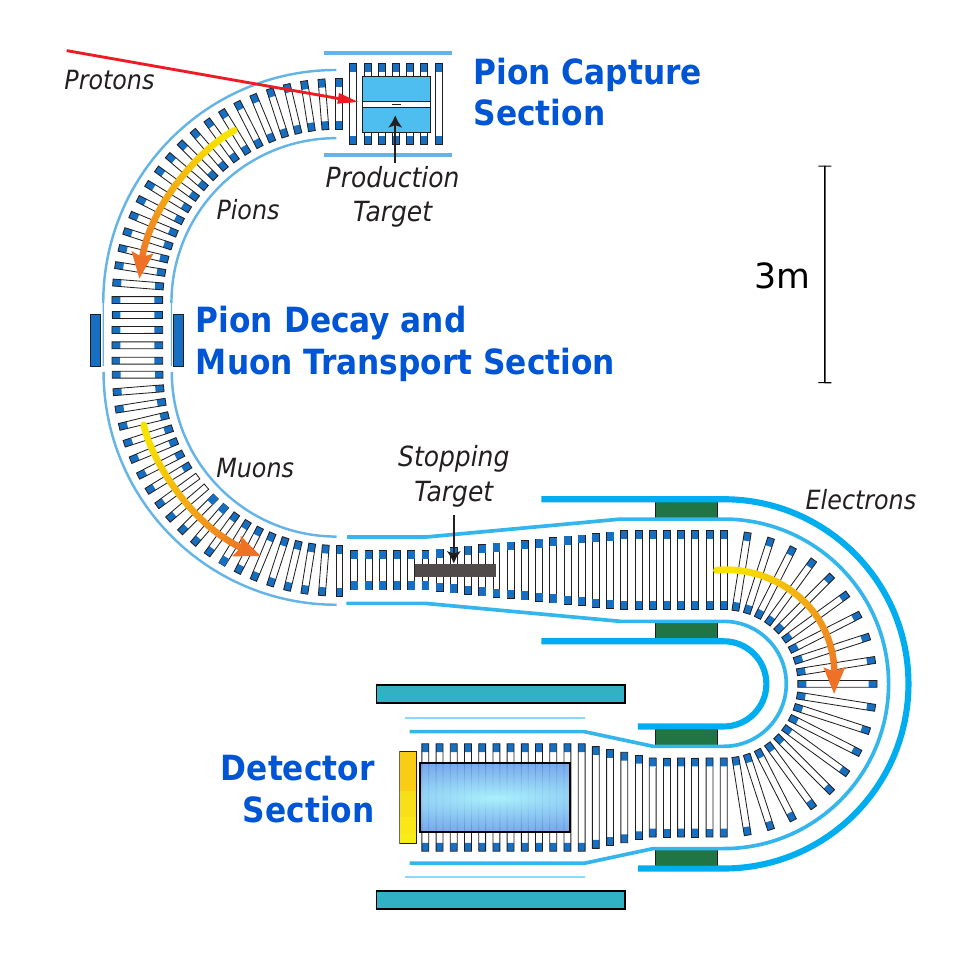}%
\caption{\figlabel{PhaseII-Overview} %
The \phaseII beamline configuration.  
Protons enter from the top-left, impinging on the production target towards the top-middle.
Pions produced in the backwards direction are then captured by a 5~T field and transported into the bent solenoid sections.
The bent solenoids select for low energy muons produced primarily from pion decays.
This secondary beam is then directed on to the stopping target (centre), which is made of aluminium.
Electrons produced from \mueconv are then transported to the detector along a second bent solenoid to remove other backgrounds like muon decay-in-orbit.
See text for more details.
}
\end{figure}
}

\newcommand{\FigPhaseI}[2]{
\begin{figure}[#1]
\includegraphics[#2]{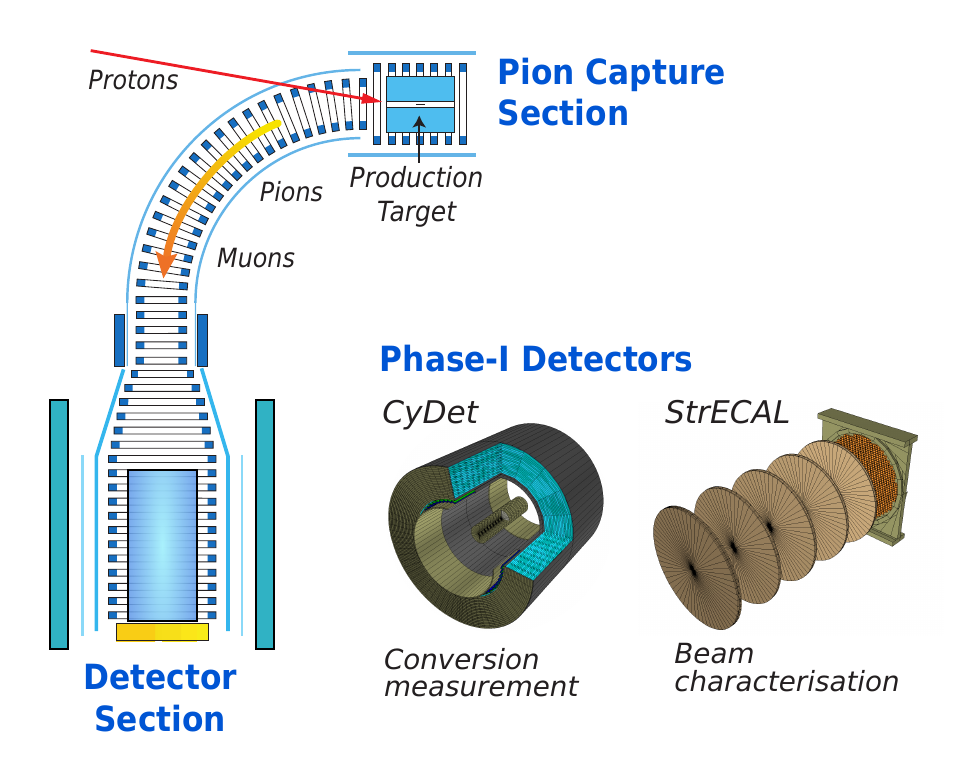}%
\caption{\figlabel{PhaseI-Overview} %
The \phaseI beamline configuration, which is identical to \phaseII up to the first 90\degree of the bent muon transport solenoid.
From there, a matching section smooths the field into the detector solenoid that will also be reused for \phaseII.
Given the two, orthogonal goals for \phaseI, two different detector systems will be used, in dedicated runs.
See text for more details.
}
\end{figure}
}

\newcommand{\FigEventDisplay}[2]{
\begin{figure}[#1]
\subfloat[][\figlabel{StrECAL-eventDisplay}StrECAL]{%
	\includegraphics[#2,trim=0 0 13cm 0,clip=true]{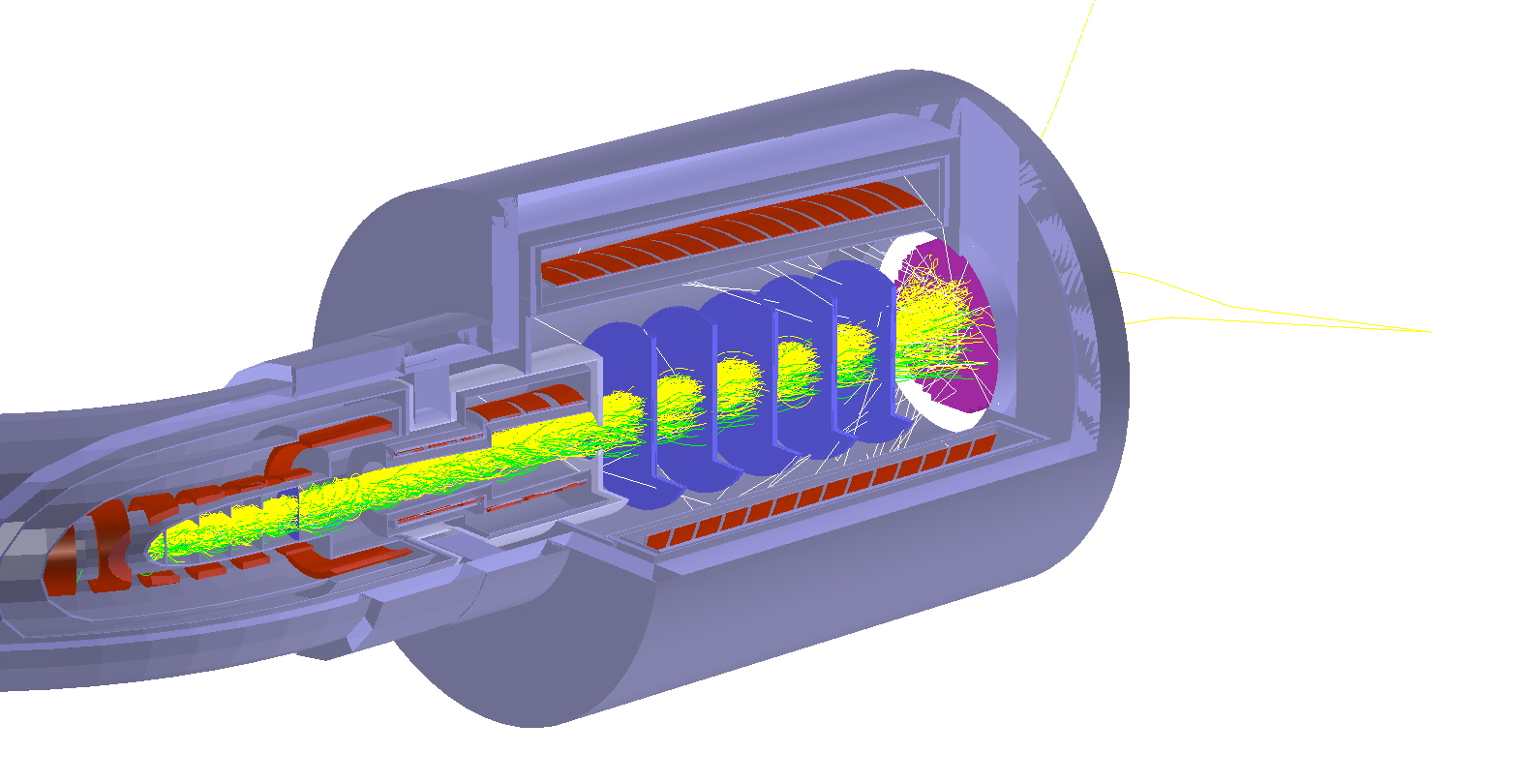}%
} \hspace{2ex}
\subfloat[][\figlabel{CyDet-eventDisplay}CyDet]{%
	\includegraphics[#2,trim=3cm 0 9cm 0,clip=true]{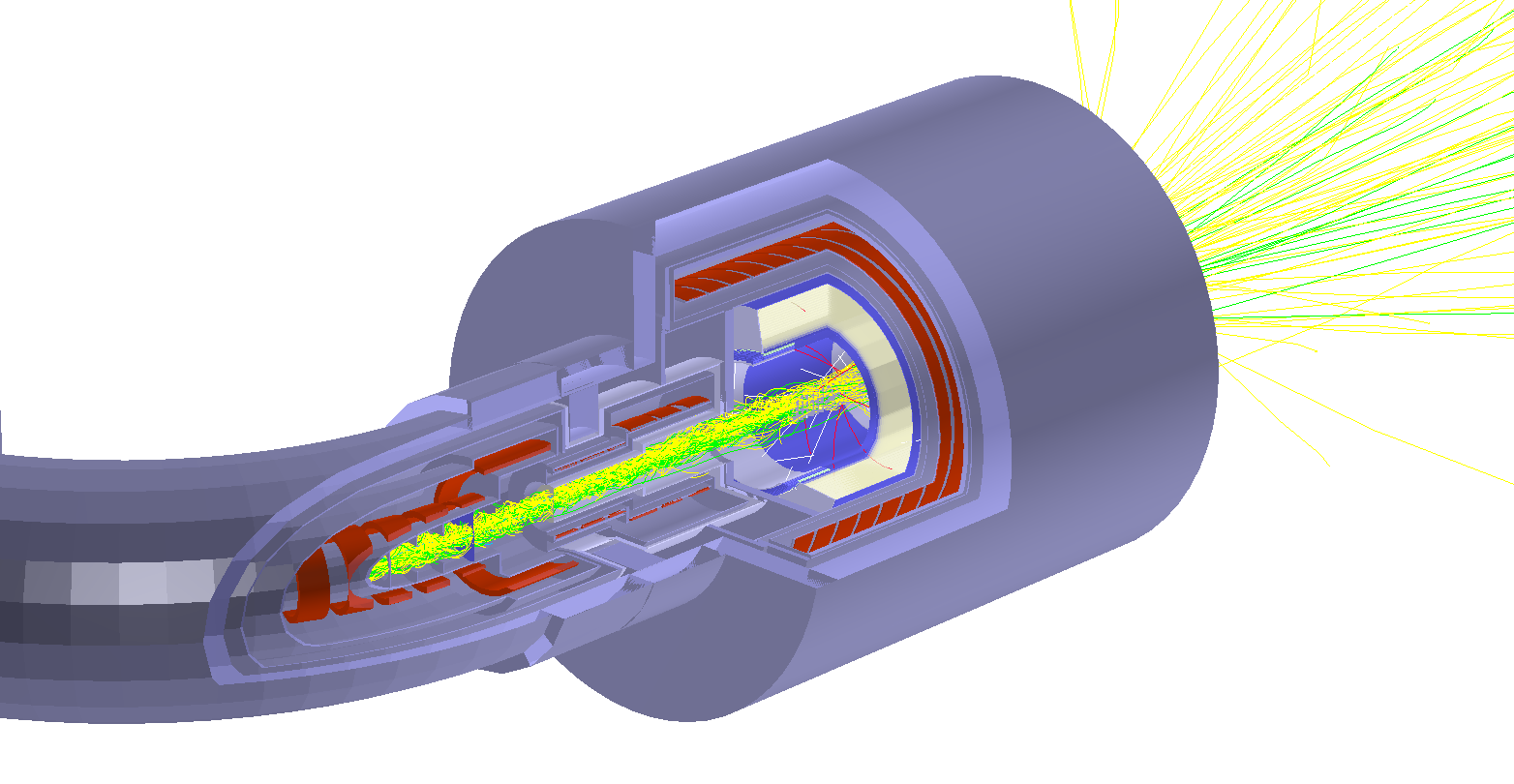}%
}
\caption{\figlabel{eventDisplays} %
Simulated event displays for the two different \phaseI detectors.  
Red tracks in the centre of the CDC are protons coming from muon capture in the target.
Track colour shows the particle ID: electrons (yellow), muons (green), photons (white).
}
\end{figure}
}

\newcommand{\FigStrECALProduction}[1]{
\begin{figure}[#1]
\subfloat[][\figlabel{straw-production}Straw Tube Construction and Prototype Tracker]{%
	\includegraphics[height=0.2\textheight]{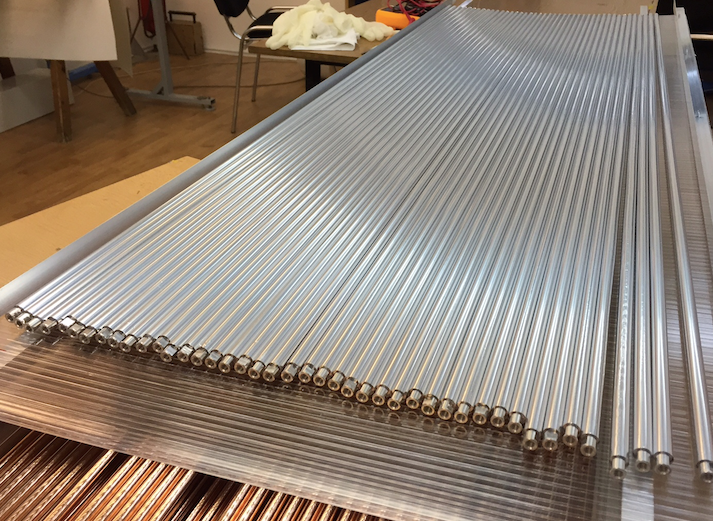}\hspace{4pt}%
	\includegraphics[height=0.2\textheight]{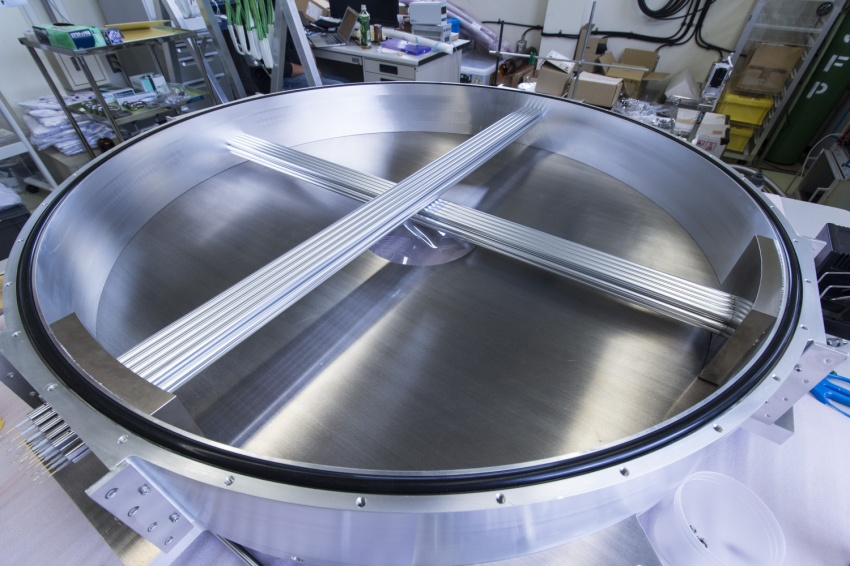}
}\\
\subfloat[][\figlabel{ecal-production}Crystal Wrapping and Test Beam]{%
	\includegraphics[height=0.19\textheight,trim=5ex 4ex 5ex 4ex,clip]{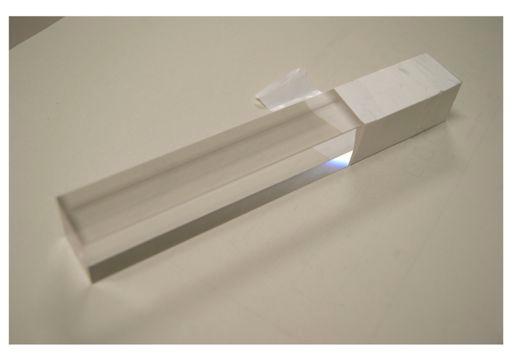}\hspace{4pt}%
	\includegraphics[height=0.19\textheight,trim=0 2cm 0 0,clip]{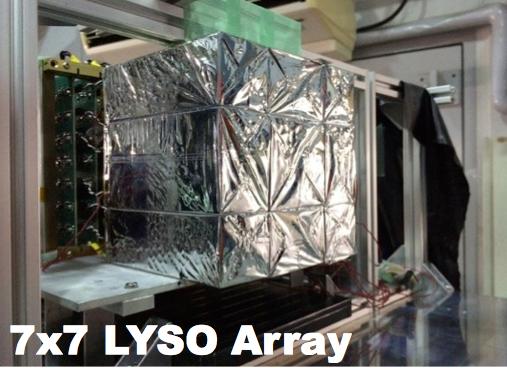}
}
\caption{\figlabel{StrECAL-production} %
Preparation of various components of the StrECAL detector.
Top: Production of 2500 aluminised-Mylar has been completed and a prototype detector has been prepared for beam and cosmic tests.
Bottom: Crystals for the ECAL wrapped and mounted for characterisation and resolution studies in an electron beam.
}
\end{figure}
}

\newcommand{\FigCyDetProduction}[1]{
\begin{figure}[#1]
\includegraphics[height=0.3\textheight]{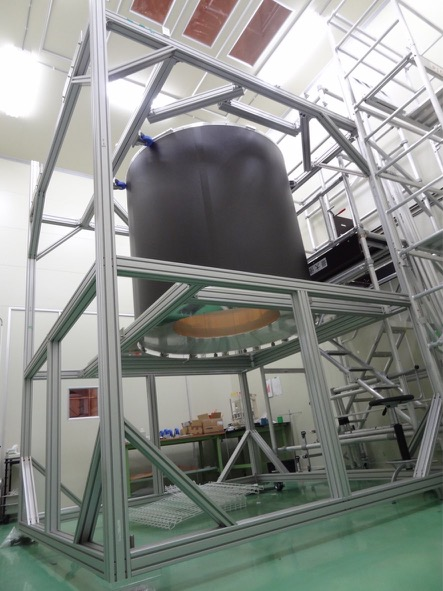}
\includegraphics[height=0.3\textheight]{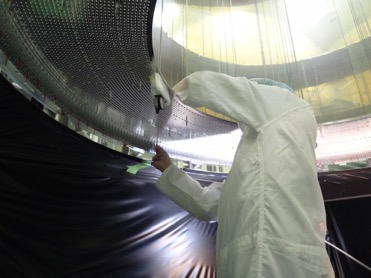}
\caption{\figlabel{CyDet-production} %
Preparation of various components of the CyDet detector.
The outer and inner walls have been purchased (left) and wire stringing has begun (right).
}
\end{figure}
}

\newcommand{\FigBuildingConstruction}[1]{
\begin{figure}[#1]
\centering
\subfloat[][\figlabel{jan}January '15]{\includegraphics[height=0.3\textheight]{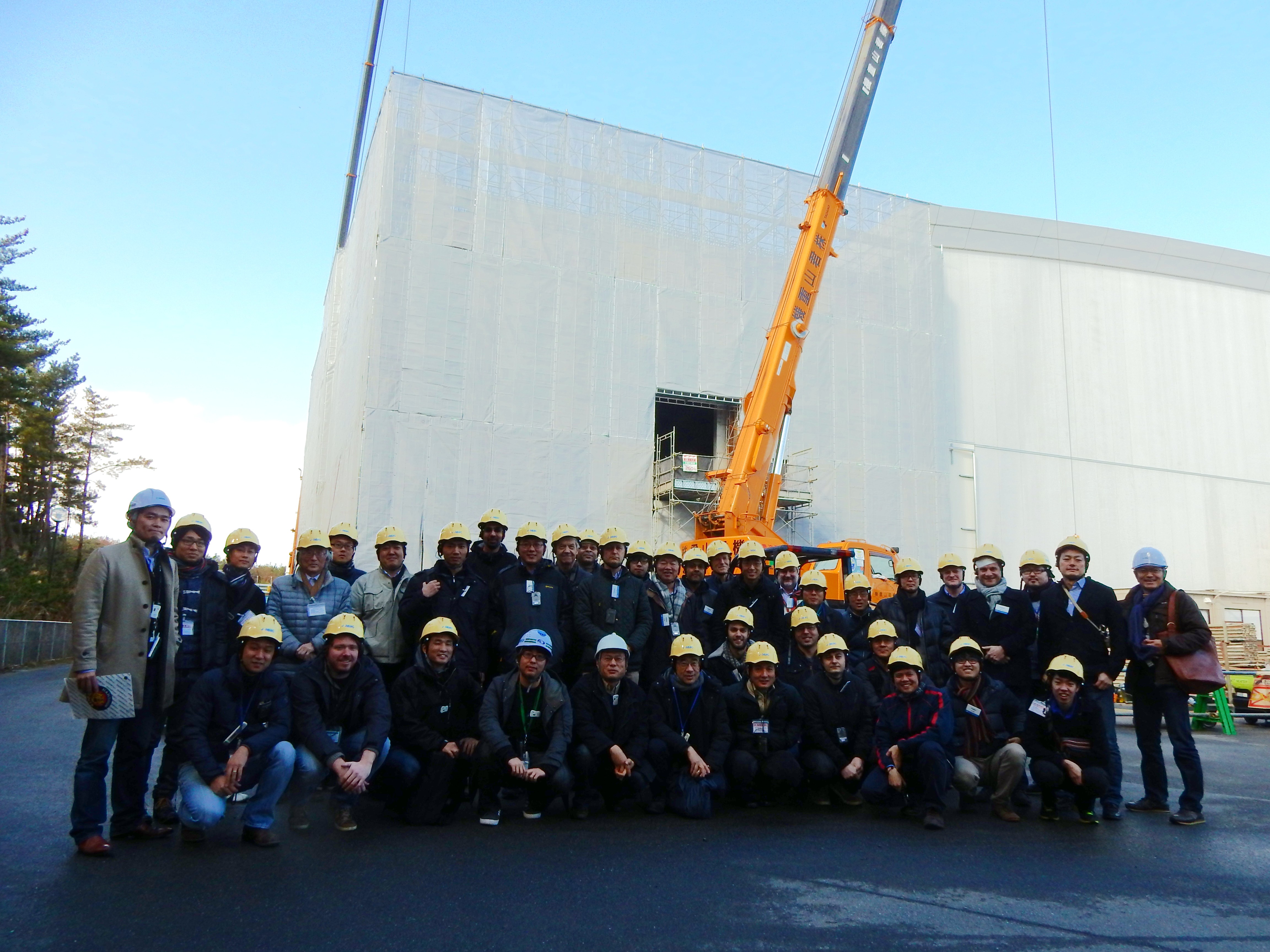}}~%
\subfloat[][\figlabel{march}March '15]{\includegraphics[height=0.3\textheight]{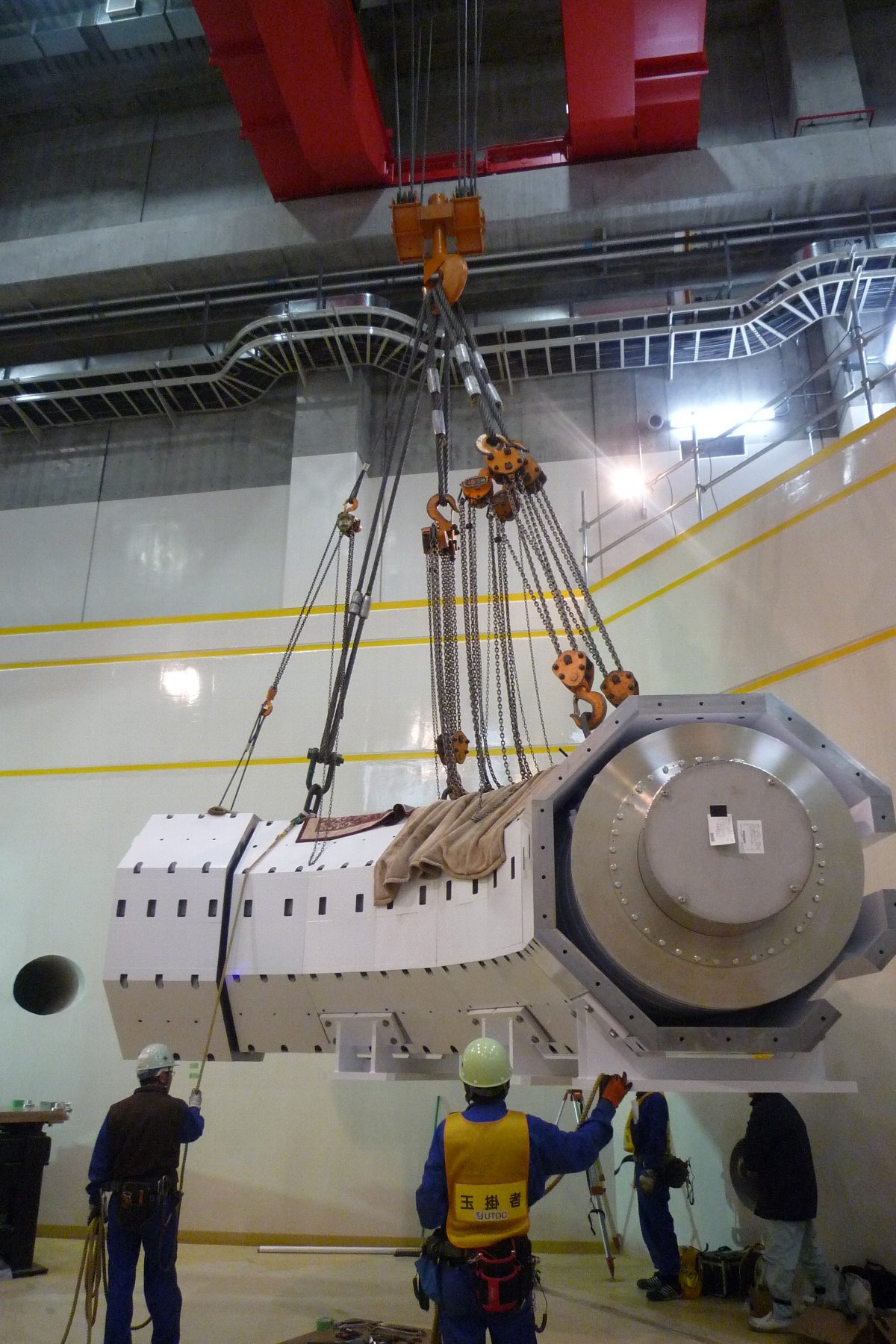}}
\caption{\figlabel{building}
Construction of the COMET building and beamline.  (\fig{jan}) The COMET collaboration in front of the nearly finished COMET Experiment Building at the 15th collaboration meeting in January 2015.
(\fig{march}) Installation of the first 90~\degree of the bent muon transport solenoid in March 2015.
}
\end{figure}
}

\newcommand{\FigDIOGeantVsCzarnecki}[2]{
\begin{figure}[#1]
\includegraphics[#2,trim=1.4cm 0.0cm 1.0cm 0.6cm,clip=true]{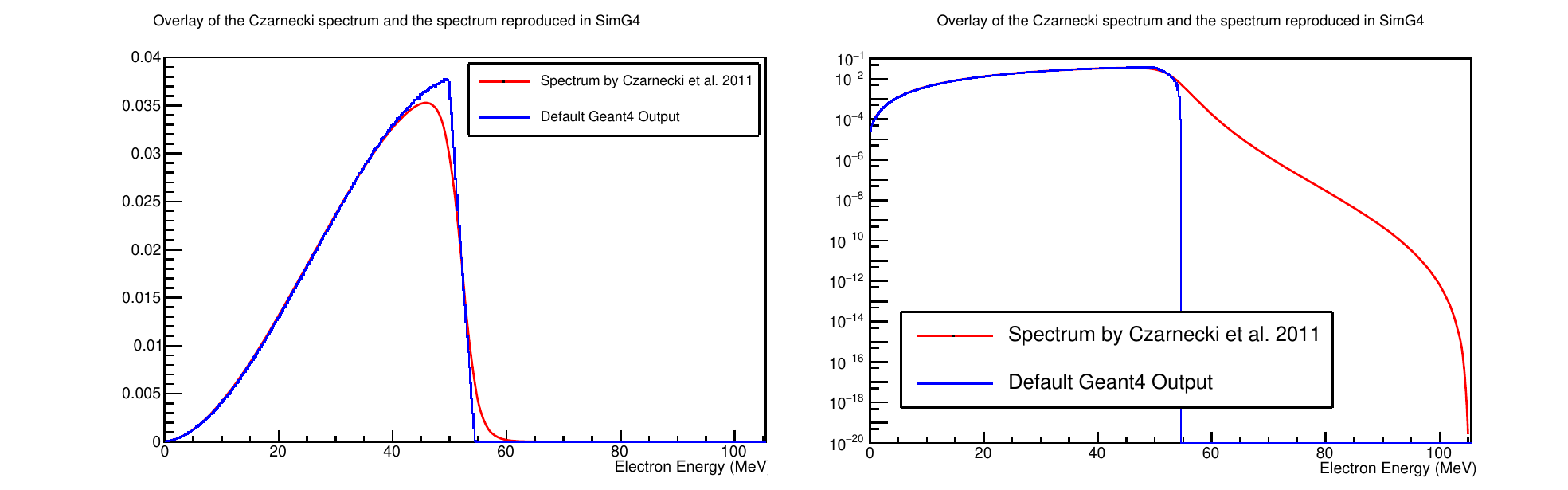}%
\caption{\figlabel{DIO-Geant4-Czarnecki} %
	Comparison of the energy spectrum of the electron emitted from bound muon decay in aluminium as predicted by theory \cite{czarnecki2011}(red) and as modelled in standard Geant4 (blue).
	The plot on the left uses a linear scale, where it is clear that the Geant4 distribution is very similar to the free muon decay, whilst the plot on the right uses a log scale
	where it can be seen that the tail of the bound muon decay, although falling very steeply, does extend up to the \mueconv signal energy of 105~MeV.
}
\end{figure}
}

\newcommand{\FigPhaseIIFieldmap}[2]{
\begin{figure}[#1]
\includegraphics[#2,trim=0 0.3cm 0 2.6cm,clip=true]{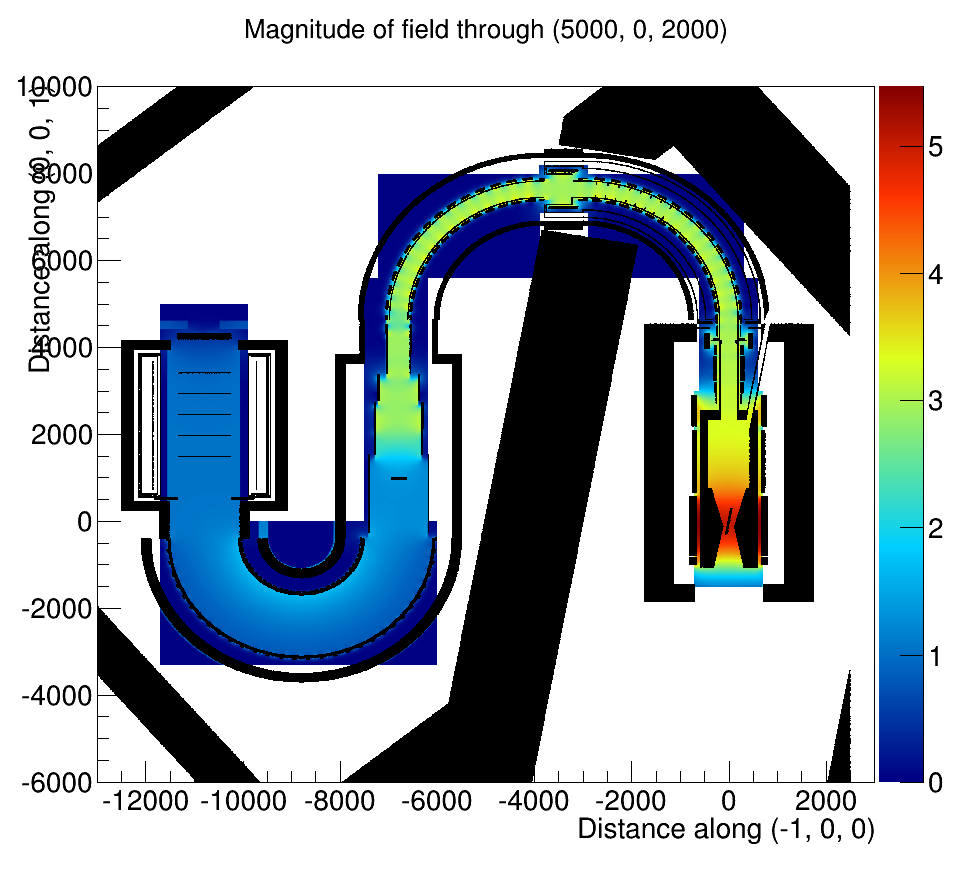}%
\caption{\figlabel{PhaseII-fieldmap} %
The recently updated \phaseII fieldmap calculation overlaid with the geometry of the Phase-II beamline and experiment hall.
}
\end{figure}
}

\newcommand{\FigGeometryUpdate}[2]{
\begin{figure}[b]
	\subfloat[][\figlabel{phaseI}\phaseI]{\includegraphics[height=0.25\textheight]{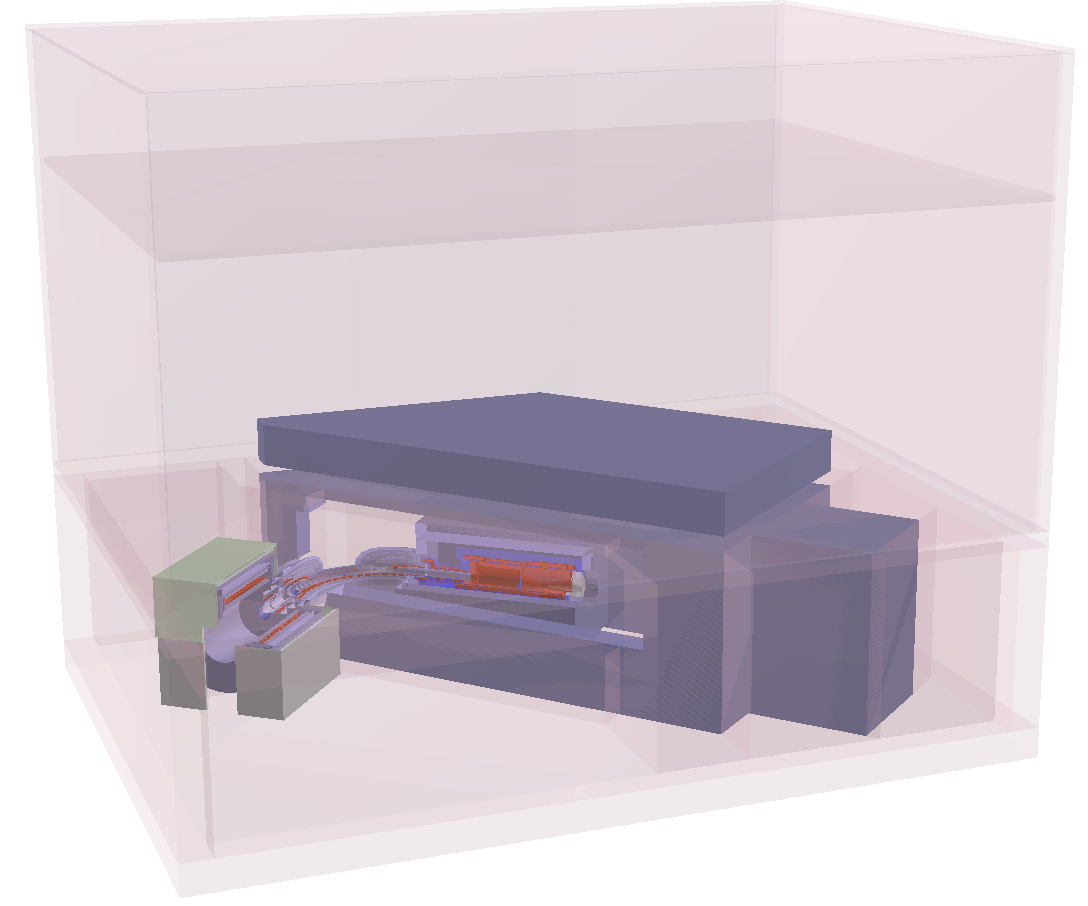}}\hspace{2ex}%
\subfloat[][\figlabel{phaseII}\phaseII]{\includegraphics[height=0.25\textheight]{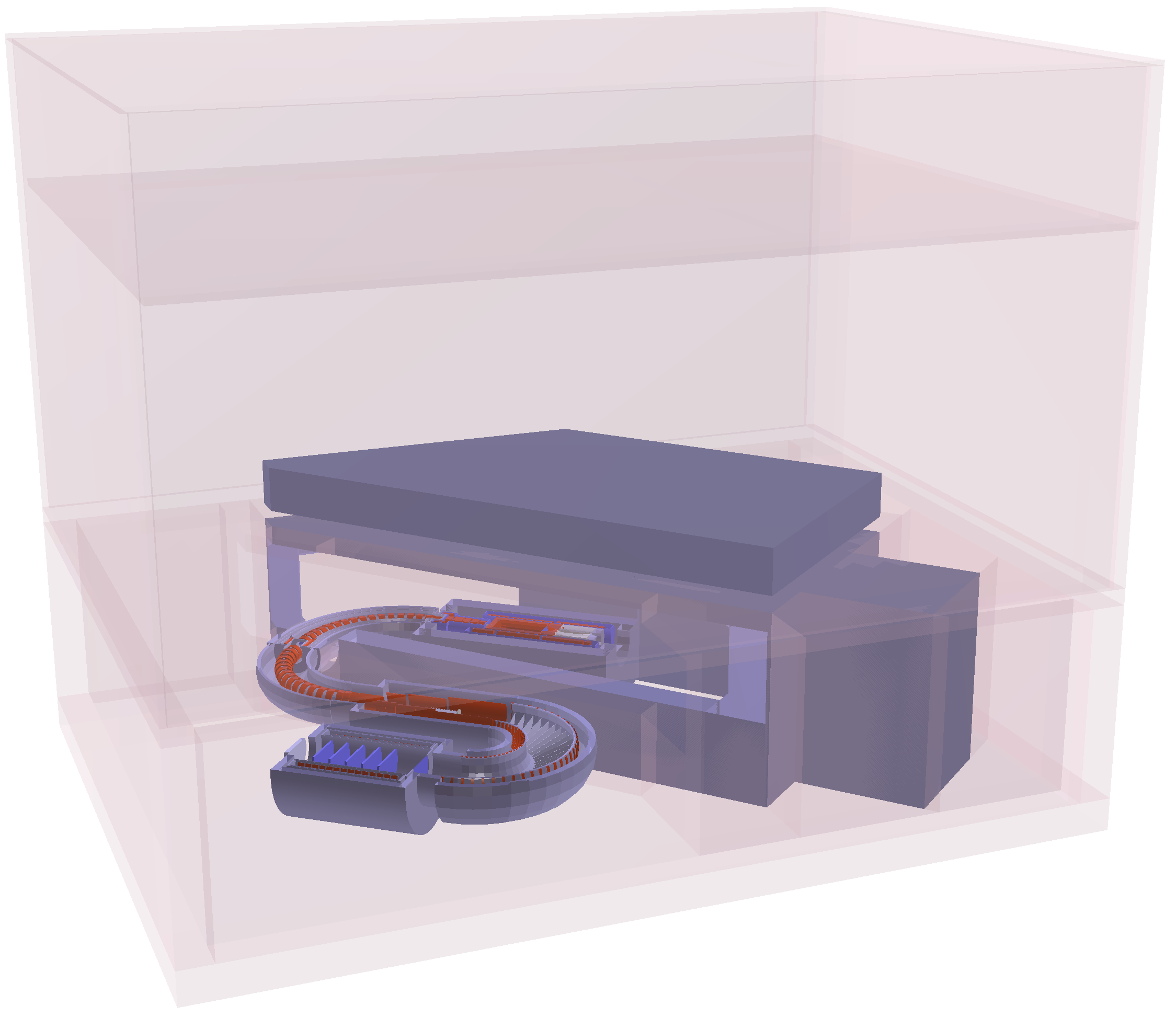}}
\caption{\figlabel{update-experimentHall} %
The current representation of the geometry for both phases
in the COMET simulation, which has been updated significantly since the NuFact conference.
}
\end{figure}
}

\newcommand{\FigBeamFlux}[2]{
\begin{figure}[#1]
	\subfloat[][\figlabel{schematic}]{\includegraphics[height=0.25\textheight,trim=0.5cm 0.5cm 0 -1cm,clip]{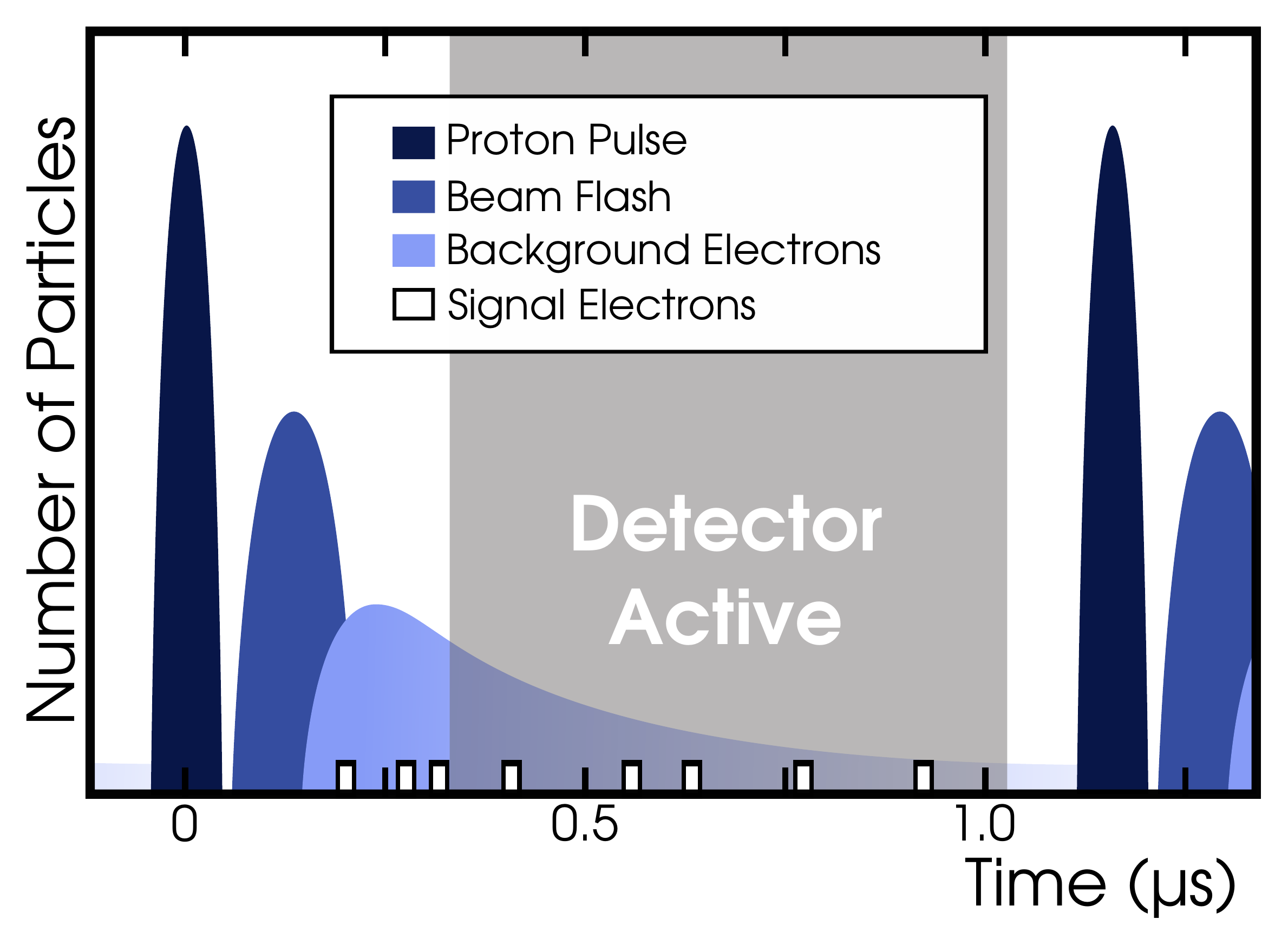}}\hspace{1cm}
	\subfloat[][\figlabel{beam-flux}]{\includegraphics[height=0.25\textheight,trim=0.5cm -0.3cm 0 0,clip]{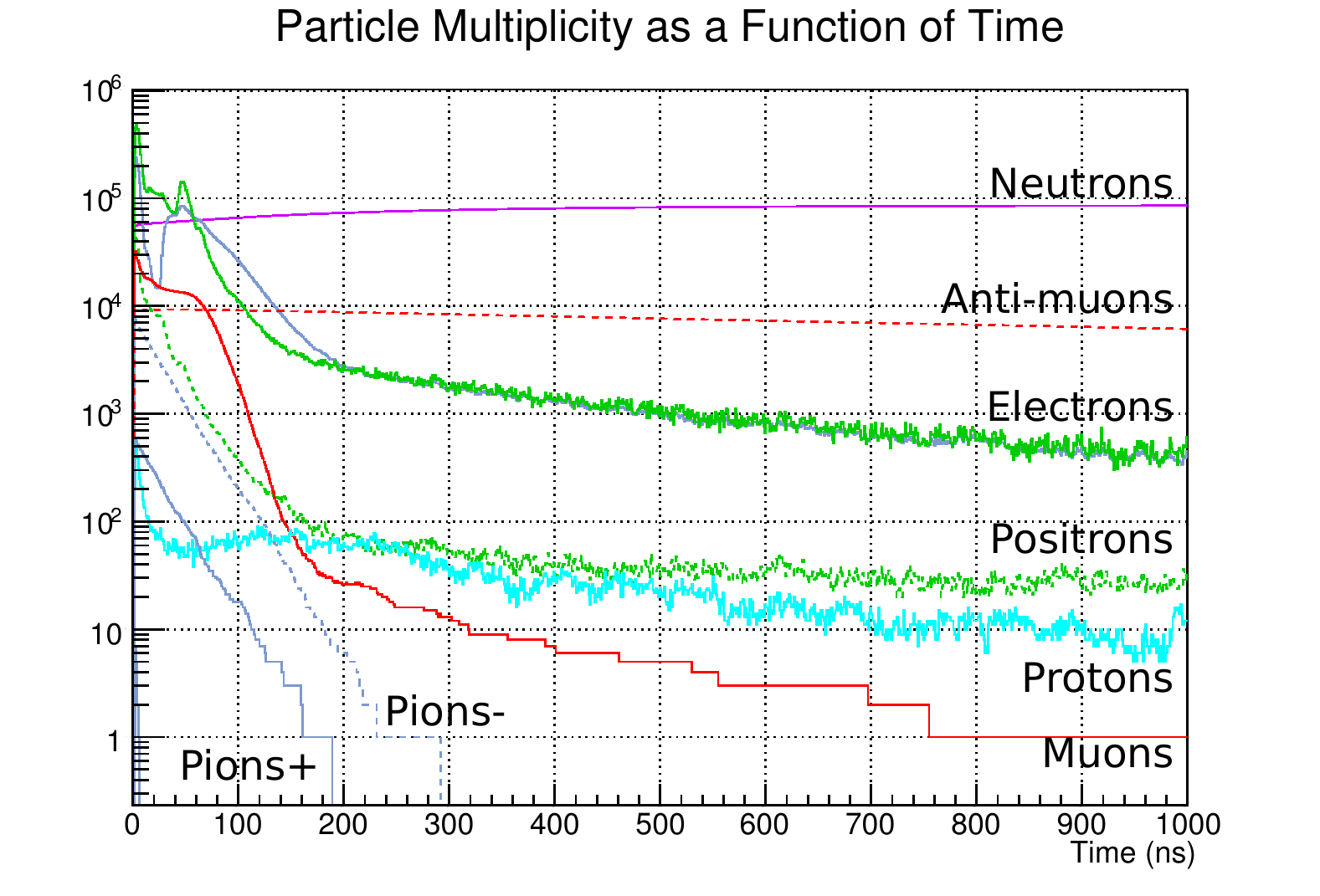}}
\caption{\figlabel{timing} %
Timing structures in COMET.
\protect\subref{fig:schematic} Schematic of the bunch structure and time-gated detector window used to reduce prompt beam-related backgrounds.
\protect\subref{fig:beam-flux} 
Simulated particle fluxes, integrated over the entire experiment and produced by protons at $t=0$. 
The y-axis is in arbitrary units, but normalised to give the correct relative flux for each particle type.
}
\end{figure}
}

\begin{document}


\title{An Overview of the COMET Experiment and Its Recent Progress}
\thanks{\it Presented at NuFact15, 10-15 Aug 2015, Rio de Janeiro, 
Brazil [C15-08-10.2]}



\author{Benjamin E. Krikler}
\email[]{benjamin.krikler07@imperial.ac.uk}
\thanks{On behalf of the COMET collaboration}
\affiliation{Imperial College London}


\date{\today}

\begin{abstract}
Forbidden in the Standard Model, Charged Lepton Flavour Violation is a strong probe for New Physics.
The COMET Experiment will measure one of these processes: that of COherent Muon to Electron Transitions, 
where a muon converts to an electron in the presence of a nucleus without the emission of any neutrinos.
COMET aims to improve the current limit on this process  by four orders of magnitude.
Being built in two phases at J-PARC, Tokai, Japan, COMET will first take data in 2018, where it should
achieve a factor 100 improvement. 
This report gives an overview of \mueconv and the COMET experiment as well as a summary of the recent progress
in construction and design.
\end{abstract}


\maketitle

\section{Introduction}
Lepton flavour conservation has been a key ingredient in our description of the world of particle physics since the first experiments showed a muon to decay to an electron only if accompanied by the emission of two other massless (or so they thought) fermions.
Tests of the validity of this conservation have continued, through searches for neutrinoless muon decay to an electron accompanied by either a photon, an electron-positron pair, or in the presence of an atomic nucleus.

This report is concerned with the last of these processes, muon to electron conversion, where a muon stopped around an atomic nucleus converts to an electron without the emission of any neutrinos.
In particular, the COMET experiment will search for COherent Muon to Electron Transitions, where the nucleus is left unchanged.
COMET is being built at the Japanese Proton Accelerator Research Centre (J-PARC), in Tokai, Japan and will first take data in 2018, during \phaseI.
\phaseII shall follow at the beginning of the next decade, and aims to improve the sensitivity to the \mueconv process by four orders of magnitude compared to the present limit.
The following sections gives an overview of muon-to-electron conversion, an outline of the COMET experiment as a whole, and a summary of the recent progress
in its construction and design.

\section{The Muon-to-electron conversion Process}
\FigFeynmanDiagrams{t}
Muon-to-electron conversion occurs as the neutrinoless decay of a muon in the presence of an atomic nucleus.
Since no neutrinos are emitted, and if the nucleus is left unchanged, the process is essentially a two-body interaction such that the energy of the out-going electron 
has a fixed value, $E_e$, given by the equation:
\begin{equation}
	E_e = M_\mu - E_\textrm{Binding} - E_\textrm{Recoil}
\end{equation}
where $M_\mu$ is the mass of the muon, $E_\textrm{Binding}$ is the binding energy of the original muon-nucleus system, and $E_\textrm{Recoil}$ is the recoil energy of the nucleus.
The last two terms are small compared to the muon mass, so that the \mueconv signal occurs at close to the value of the muon mass and is well separated from electrons of Standard Model muon decay (with neutrino emission), which for a free muon
can only achieve energies up to half the muon mass.  

The fact that the signal for \mueconv is a single, mono-energetic electron that is well separated from any intrinsic physics background makes it a very experimentally attractive search.
Further background suppression can be achieved using timing information of the process, which is fixed to the lifetime of the muonic atom.
In aluminium, the target of choice for COMET, the lifetime of the muon is about 864~ns, whilst the signal energy is $E_e=104.97$~MeV.

Typically for muon-to-electron conversion one defines the rate of such a process by normalising to the rate of muon nuclear capture, since this cancels the theoretical uncertainty on the muon wave-function.
The current limit on \mueconv comes from the \sindrumII experiment \cite{sindrum2006}, which used a gold target and set a 90\% confidence limit on the Conversion Rate (C.R.) at \senseSindrum.
COMET will be built in a staged approach hoping to improve this limit (but using an aluminium target) by about two orders of magnitude at each stage.  
The Single-Event-Sensitivity (SES) is the figure of merit for a \mueconv experiment's ability to observe the signal process.
It is equivalent to the minimum value of the conversion rate where the experiment can still expect to see one signal event during the run.
For COMET \phaseI, our SES is \sensePI, which should improve to \sensePII for \phaseII.

Constraining ourselves to muon-to-electron conversion where the nucleus is left unchanged causes coherent terms to dominate the interaction with the nucleus, which should provide
a significant enhancement to the rate, compared to the incoherent form.  
Yet whilst this fixes the value of the signal energy to be independent of the underlying physics the exact rate is highly model dependent.  
In principle, neutrino oscillations alone produce this sort of charged lepton flavour violation through penguin diagrams like that shown in \fig{FD-dipole}.
However, if this were the only mechanism, the process would be highly GIM suppressed by the tiny mass squared difference of the neutrinos to conversion rates of order $\mathcal{O}(10^{-54})$.
As a corollary, if New Physics is to be seen it must be well beyond both the Standard Model and even neutrino oscillations.
There is no dearth however of models that give measurable conversion rates, including leptoquarks, $Z$-primes, extended Higgs couplings, supersymmetry, and of course heavy neutrinos \cite{Altmannshofer2009ne}.


\section{The COMET Experiment}
Maximising signal acceptance whilst minimizing background is an important aspect of any experiment, and COMET is no exception, given the rarity of the signal process
and challenge of providing an intense, low-energy muon source.
The key backgrounds that must be optimised against include intrinsic ones which occur when negative muons stop in a target, processes related to impurities in the beam,
detector effects such as particle misidentification and pile-up, and cosmic backgrounds.

Beam related backgrounds are predominantly removed using a pulsed proton beam in combination with a delayed-time detector window, show schematically in \fig{schematic}.
For COMET, protons arrive at the primary production target in 100~ns bunches.  
This, convoluted with the transport of the secondary pion and muon beam, leads to a beam flash at the stopping target of roughly 200~ns at \phaseI, as shown in \fig{beam-flux}.
Since the muon lifetime in aluminium is 864~ns, most of the beam flash can be ignored with only a small reduction to the signal acceptance by using a delayed-time window beginning around 500 to 700~ns, with additional delay to account for detector response time and occupancy.
This puts two additional requirements on the beam timing: that the bunch spacing be large compared to the muon lifetime in the target, and that the number of protons reaching the production target in between bunches (quantified by the extinction factor, the ratio of protons between bunches compared to the number in a bunch) be sufficiently low.
On the first requirement, the main ring at J-PARC has a bucket separation of about 550~ns \cite{TDR2014}.
For COMET, only every other bucket will be filled leading to a bunch separation of about 1.17~$\mu$s.
For the second requirement of few protons arriving out-of-time, extinction factors of around $10^9$ were thought necessary, whilst measurements in late 2014 demonstrated that extinction levels at around $10^{12}$ were achievable using the J-PARC Main Ring injection kicker in a novel double-kick mode \cite{TDR2014}.
\FigBeamFlux{t}{width=0.4\textwidth}

The intrinsic backgrounds for COMET are any process involving a negative muon stopping in the target and resulting in an electron close to 105~MeV.
This includes radiative muon capture followed by pair-production from the photon (either internally or externally) and the Standard Model decay of the bound muon.
Of the two, it is the bound decay which is expected to contribute most to the
background rate, since in aluminium the relative rate for muon decay to
the radiative form is around $\mathcal{O}(10^{-5})$ \cite{Measday2001243},
which is further suppressed by the need for the photon to then asymmetrically
pair-produce.
Whilst electrons from the free decay of the muon cannot be produced above half the muon mass (in the muon rest frame), once bound to a nucleus, the end-point of the decay shifts up to the \mueconv signal energy, since the neutrinos can now be configured to carry away vanishingly little energy.  
Clearly the probability for such an arrangement is similarly vanishingly small, and indeed the spectrum falls very steeply above 55~MeV.
Nonetheless, distinguishing electrons of this process from true \mueconv electrons can only be achieved with sufficiently high electron energy resolution.
This implies detectors with low-material budget but also that the stopping target be optimised to reduce scattering of electrons as they leave it.
Ideally then, one would use as a thin a target as possible, but this must be offset by the decrease in the ability to stop the muon beam, which favours a thicker target.
The need for a low energy muon beam therefore arises, in order to reduce the muon stopping distance.
A similar constraint comes from the need to keep high-energy beam electrons out of the detector, although the pulsed beam and delayed-time detector gate also help suppress such potential backgrounds.

The muon beam must therefore have very high intensity, be of low energy, and contain few impurities.
The COMET experiment achieves these properties with two novel approaches:
capturing backwards emitted pions and muons from the production target using superconducting solenoid fields,
and a combination of bent solenoids, vertical dipole fields and collimators along the muon beam transport.
Both of these remove the high-energy components of the beam whilst maintaining a high muon intensity, and the long decay length of the bent transport solenoids additionally improves beam purity since most pions will decay.
\FigPhaseII{t}{width=0.5\textwidth}

The transport dynamics of bent solenoid fields are complex, but to first order one can consider the field as modifying the simple helical trajectory of a particle moving through a straight solenoidal field.
The change to this behaviour is produced by two effects:
a) the fact that the field is no longer uniform in the transverse plane induces a ``grad-B'' drift proportional to the transverse momentum, and
b) the coordinate system for the centre of the helix is now cylindrical (since it still follows the field lines, which themselves are cylindrical), introducing a centrifugal pseudo-force, proportional to the longitudinal momentum.
Both of these effects result in a drift of the centre of a particle's helix in the vertical direction (out of the plane of bending) by a height, $D$, given by the equation:
\begin{eqnarray}
	D=&\frac{1}{qB}\frac{S}{R}\Big(\frac{p^2_l+\frac{1}{2}p^2_t}{p_l}\Big)\\
	\propto&\frac{p}{q}\big(\cos(\theta)+\frac{1}{\cos(\theta)}\big)
\end{eqnarray}
where $p$, $p_l$ and $p_t$ are the total, longitudinal and transverse components of momentum respectively, $\theta$ the pitch angle of the helical particle trajectory, $S$ is the length travelled along the bent solenoid, $R$ is the radius of bending for the solenoid, $B$ the magnetic field strength along the solenoidal axis, and $q$ the charge of the particle being transported.
Since the dependence on the pitch angle is weak for angles up to around 60\degree, the dispersion introduced by this drift is mostly governed by the magnetic rigidity ($p/q$) of the particle.
By superposing a dipole field, which causes a drift that is independent of the particles momentum, one can tune the beam so that a nominal momentum (and pitch angle) is kept on axis,
whilst particles with other rigidities are removed by the beam pipe and carefully tuned collimators.

\FigPhaseI{t}{width=0.5\textwidth}
\subsection{COMET \phaseII}
\Fig{PhaseII-Overview} shows a schematic of the configuration for \phaseII.  
From this it can be seen that the muon transport beamline captures pions coming backwards from the production with respect to the proton beam, which itself enters from the top-left corner of the image.
In \phaseII, this secondary beam is then transported around 180\degree of bent solenoid (with a small straight section in the middle for possible collimators and field matching).
The beam is then directed on to the stopping target which is made of 200~$\mu$m thick aluminium disks, and followed by a beam blocker that should absorb any beam that does not stop in the target.
Electrons produced in the target are then collected by a gradated magnetic field, and transported around a second section of bent solenoid with a much larger aperture.
The dipole field along this region is tuned to remove low energy electrons from decay-in-orbit and other charged particles coming from the stopping target.
Additionally, having no line-of-sight between the target and the detector helps reduce backgrounds from neutral particles such as photons from radiative muon capture.
Finally, the electrons enter the detector system formed by a series of straw tracker planes and a crystal electromagnetic calorimeter (ECAL).

\subsection{COMET \phaseI}
\FigEventDisplay{b}{width=0.4\textwidth}
Given the number of new techniques being employed for COMET \phaseII there are many uncertainties associated with the expected production yields, beamline dynamics and consequently the final background rates.
It was therefore decided to take a staged approach, such that COMET \phaseI will build the production target and first 90\degree of muon transport beamline.
\phaseI will operate two different configurations, either to perform beam
characterisation or to make a \mueconv measurement, aiming for a hundred-fold improvement on the limits set by the \sindrumII experiment.
\Fig{PhaseI-Overview} illustrates the beamline that will be used for \phaseI, showing how the first 90\degree for \phaseII will be built at this stage.
Whilst the \phaseII detector would be suitable for characterising the beam, given the reduced distance between production target, stopping target and detector in \phaseI, it would not be suitable for performing the physics measurement.
As such, a second system using a cylindrical detector (CyDet) has been developed for the \phaseI physics measurement, whilst the straw tube tracker and crystal electromagnetic calorimeter that make up the \phaseII detector are constructed in parallel.
\Fig{eventDisplays} shows simulated event displays for the two different \phaseI detectors.

\section{Construction and On-going Research and Development}
\FigBuildingConstruction{h}
\subsection{Facility Construction}
\Fig{building} shows two images of the construction of the COMET facility.  
The COMET building joins on to the existing Hadron Hall at J-PARC and contains the experiment area on the lowest level, a staging and craning area on the first and second floor, and offices and control rooms on the top floor.
In January 2015, the fifteenth COMET collaboration meeting was held in J-PARC and KEK where the collaboration were able to view the nearly finished building.
By March 2015, the first beamline components were being installed, starting with the first 90\degree of bent muon transport solenoid that will be used for \phaseI.

\subsection{The CyDet: a Cylindrical Detector for the \phaseI Physics Measurement}%
\FigCyDetProduction{b}%
In order to achieve the desired two orders of magnitude improvement over \sindrumII at \phaseI, the detector needs to be blind to most of the beam flash, and the large 
number of low-energy electrons produced by bound muon decay in the target.
The detector known as the CyDet (Cylindrical Detectory) will be tasked with this measurement.
It combines a Cylindrical Drift Chamber (CDC) with two rings of Cherenkov Hodoscope and Scintillation counters which provide a trigger and $t_0$ for each event.

Contained in a co-axial solenoidal field, the inner radius of the CDC is tuned such that the detector is blind to most of the beam flash, which will enter and remain in the region close to the solenoidal axis given its relatively low momentum.  
The same is true for the bulk of the bound muon decay spectrum, the majority of which has momentum less than 60~MeV.
Given the stopping target diameter and the 1~T field magnitude in the detector, the inner wall of the CDC is set to 60~cm which means electrons with less than 60~MeV coming from the target are unable to reach the detector.
\Fig{CyDet-eventDisplay} shows an event display from a simulation of the CyDet with the beam flash from a single proton bunch passing through it.

The CDC will contain 20 layers of all-stereoscopic wires, each layer essentially twisted in the opposite direction to the last with a stereo angle of $\pm4\degree$.
On each layer there are between 800 and 1200 wires, including sense wires that will be made of 25~$\mu$m gold plated tungsten, and 120~$\mu$m pure aluminium field wires.

Construction of the CDC is well under way with about 40\% of all wires already strung.
In total 150~days are expected to be needed to complete wire stringing, which should finish in November when tensioning checks can be performed before transportation to the COMET facility.
In the meantime, tests using prototype versions have been on-going using both cosmic rays and electron beams at Tohoku University in Japan.
Additionally, significant work is under way to study the use of a purely track-based trigger, which could allow the triggering hodoscopes to be removed and higher beam rates supported.

\subsection{The StrECAL: a Straw Tube Tracker and Crystal ECAL Detector for \phaseI Beam Measurements and Prototyping for \phaseII}%
\FigStrECALProduction{b}%
The StrECAL, consisting of several Straw Tube Tracker stations (five in \phaseI, but possibly more for \phaseII) followed by an ECAL is able to measure both momentum and energy of particles
over the full cross section of the beam as is shown in \fig{StrECAL-eventDisplay}.
As such, in \phaseI it will predominantly be used to profile the beam and understand the impurity rate, momentum distributions, transport beam optics, and additionally the production target distributions.
Whilst it may additionally be used for a physics measurement, it is expected to be less sensitive than the CDC given that its central region will suffer high occupancy from low energy beam particles and bound muon decay electrons.
Additionally, by building, testing and running the StrECAL in \phaseI, significant understanding of the detector can be progressed in anticipation of \phaseII.

For the Straw Tube Tracker, straw production for \phaseI was recently completed, with 2500 tubes being made in total.
The straws use the single-seam welding procedure developed for NA62~\cite{NA62-Hahn:1404985} which is able to reduce the straw thickness whilst maintaining mechanical strength.
In \phaseI straws of 9.8~mm diameter will be made from 20~$\mu$m aluminised Mylar.
For \phaseII there is some scope to reduce these parameters further to 5~mm straws made from 12~$\mu$m Al-Mylar tape.
Wires of gold-coated tungsten of 25~$\mu$m diameter will be used.

In parallel, work on the ECAL is well under way.  
The decision to use LYSO (Lutetium Yttrium Sulphate) was taken in February 2015 and procurement has already begun with 200 crystals expected to have been purchased by the end of the fiscal year.
Despite its increased cost, LYSO was chosen over GSO due to its increased light yield and response time which lead to an improved energy resolution and greater robustness against pile-up.
These properties have been tested and confirmed in dedicated beam tests at PSI, Zurich and at Tohoku University in Japan.
Each crystal is $2\times2\times12$~cm which is about 10 radiation lengths for an electron at 105~MeV and in total, by \phaseII, about 2272 crystals will be used.
Avalanche photodiodes (APDs) will be mounted to each crystal which will then be wrapped in Teflon tape and grouped into modules of $2\times2$ crystals to be wrapped in aluminised Mylar.

As well as dedicated beam tests to check the intrinsic crystal resolution, beam tests have also taken place to check the light collection efficiency for different APD sizes, calibration techniques and Particle Identification (PID) methods that use just the ECAL without additional information from the Straw Tracker.
Additionally, work on the read-out and trigger system has progressed with radiation tests on the key components being completed.
A scheme for in-situ calibration and temperature measurements is being developed since the light yield of the crystals and the quantum efficiency for the APDs are both temperature dependent and sensitive to local fluctuation.
Finally a test-bench specifically to characterise and measure each crystal is being prepared.

\subsection{Simulation, Offline Software and Expected Backgrounds }
\FigDIOGeantVsCzarnecki{b}{width=0.9\textwidth}
Simulating the COMET experiment is no easy task, given that some $10^{19}$ Protons are expected to be stopped in the production target at \phaseI whilst to achieve the desired sensitivity fewer than one background events should occur.
This means the simulation needs to be both highly efficient and highly detailed, with accurate modelling of the geometry and material properties, magnetic fields, and underlying physics processes.

There are two areas in COMET in particular where current knowledge is insufficient at the level needed for COMET.
The pion and muon yield from the production target with protons of 8~GeV and coming in the backwards direction is not well known, given that no experiment has operated in this regime before.
Data from the HARP experiment~\cite{HARP2007} and some input from the MuSiC \cite{MuSIC2014} experiment has been used, in addition to running with multiple different hadronic production models, including Geant4 (QGSP\_BERT\_HP) \cite{Geant42003}, MARS \cite{MARS1995}, Fluka \cite{FLUKA2005} and PHITS \cite{PHITS2002}.

The second aspect of COMET that is not well modelled is the physics around muons bound to aluminium nuclei.
For muon nuclear capture, whilst the branching fraction is known to be about 61\% for a stopped negative muon in aluminium \cite{Measday2001243}
the rates for subsequent charged or neutral particle emission are not well known.
With this goal in mind the \alcap experiment \cite{MeNufactAlCap2015} has been set up to measure these daughter particles and the results from this experiment are feeding back in to the COMET simulation.
At the same time, the theoretical spectrum for electrons from bound muon decay in aluminium has only been calculated in the last few years.  
The default Geant4 spectrum for this process does not accurately reflect this however, with only a rather crude parametrisation of the way the free muon decay spectrum is modified that leaves an abrupt cut-off at 55~MeV.
A comparison of the two spectra can be seen in \fig{DIO-Geant4-Czarnecki}.

The simulation chain begins with the production target, where any of the codes listed above can be used, then moves on to the beam and detector simulations which is based on Geant4.
Energy deposits produced at this stage can then either be fed directly into a detector and electronics response package, or via a resampling package that allows a smaller set of proton events to be robustly
resampled and smeared in time. 
The COMET Software includes all these aspects (and more) and recently reached its first stable release in April this year.
Since then, two large scale productions have taken place.

\FigGeometryUpdate{b}{height=0.3\textheight}
\section{Developments Since NuFact}
Since the NuFact conference in August, significant progress has been made on several fronts.
For the CDC, wire stringing was completed in November, and wire tensioning measurements are well under way.

Several beam test programs have been carried out including a test at PSI to look at novel particle identification (PID) methods using the ECAL, and separate Straw Tracker and CDC prototype tests at Tohoku University, Sendai, Japan.

On the simulation side, custom muon physics has been completely implemented, including the bound muon decay spectrum and preliminary results from the \alcap experiment.
Additionally, the complete experiment hall and cosmic ray veto geometry have been included as shown in \fig{update-experimentHall}.
Improvements in the calculation of the magnetic field, which can be seen in \fig{PhaseII-fieldmap} have also been included in the simulation, 
Using all of this, a third production should have started by the end of this calendar year.
\FigPhaseIIFieldmap{t}{width=0.45\textwidth}

%

\section{Summary}
The process of muon to electron conversion is a highly sensitive probe for new physics beyond the Standard Model.
The COMET experiment will make the first measurement of this process since the \sindrumII experiment in 2006.
Using several new techniques including a pulsed proton beam, backwards capture of pions and muons from the production target,
and bent solenoids combined with tuned dipole fields and collimators allow for \phaseII's four-orders-of-magnitude improvement to the signal sensitivity compared with \sindrumII.
Given the uncertainties associated with such novel approaches, COMET will first run \phaseI with a reduced transport beamline and two dedicated detectors, with data taking beginning in 
2018, and an expected sensitivity of \sensePI.  
Progress is well under way, in particular, for the building and facility construction, delivery of the first beamline sections, detector development, and offline software.

\begin{acknowledgments}
\section{Acknowledgements}
I would like to thank the COMET collaboration for giving me the opportunity to present
at NuFact and providing feedback on the presentation and this report.
Additionally I would like to thank the STFC (UK) and IoP's Research Student Conference Fund for 
funding my participation.
\end{acknowledgments}

\bibliography{record}

\end{document}